# GRS 1758-258: the first winged microquasar

Josep Martí[1], Pedro L. Luque-Escamilla[2], Valentí Bosch-Ramon[3], Josep M. Paredes[3]

[1]Departamento de Física, Escuela Politécnica Superior de Jaén, Universidad de Jaén, Campus Las Lagunillas s/n, A3, 23071 Jaén, Spain

[2]Departamento de Ingeniería Mecánica y Minera, Escuela Politécnica Superior de Jaén, Universidad de Jaén, Campus Las Lagunillas s/n, A3, 23071 Jaén, Spain

[3]Departament de Física Quàntica i Astrofísica, Institut de Ciències del Cosmos (ICCUB), Universitat de Barcelona, IEEC-UB, Martí i Franquès 1, E-08028 Barcelona, Spain

**Abstract.** The family links between radio galaxies and microquasars have been strongly strengthened thanks to a new common phenomenon: the presence of extended winged features. The first detection of such structures in a Galactic microquasar, recently reported in *Nature Communications* (http://rdcu.be/zgX8), widens the already known analogy between both kinds of outflow sources (Martí et al. 2017). This observational result also has potential implications affecting the black hole merger scenarios that contribute to the gravitational wave background.

**Introduction**

Since two decades ago relativistic jet sources are well known to share very common morphological and physical properties that broadly scale with the mass of their central compact object (Mirabel & Rodríguez 1999). The analogies between stellar-mass and supermassive black hole outflow systems have been intensively explored concerning the jet launching mechanisms in the accretion disc close vicinity. However, less attention has been paid far beyond, i.e., where the relativistic outflows strongly interact with their ambient medium. In this context, could microquasars be used as a workbench where to test the different scenarios proposed to explain the intriguing X and Z-shaped lobe morphologies of the so-called winged radio galaxies (WRGs)?

WRGs represent a non-negligible sub-class of outflow sources whose number tends to increase thanks to modern radio surveys (see e.g. Cheung 2007; Roberts et al. 2009). The physical scenarios proposed to account for the origin of their secondary lobes include, among others, spin-flip of the jet axis after the coalescence of supermassive black holes, relic jet orientations due to precession, or hydrodynamic backflow of shocked jet material. In this contribution[1], we briefly assess all these interpretations when applied to the radio lobes of the Galactic Center microquasar GRS 1758-258 where a winged Z-type morphology has been recently discovered (Martí et al. 2017).

---

[1] Presented as an invited talk to the workshop on 'Interstellar Medium and Radio Astronomy in Catalonia: the Robert Estalella Legacy', held at the University of Barcelona on 21-22 December 2017.

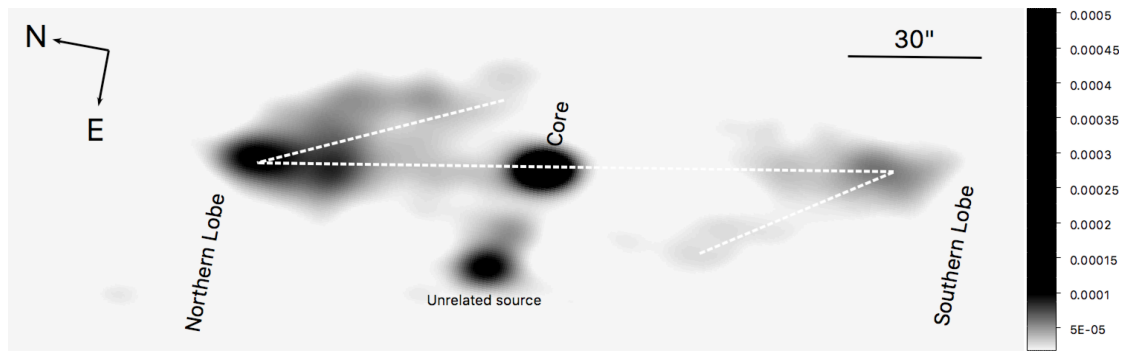

**Figure 1.** Radio map of the GRS 1758-258 jets and lobes with the Z-shaped winged features outlined by the dashed white line as an aid for the eye (using the same data as in Martí et al. 2017). The figure has been rotated counterclockwise to place the ejection axis horizontally. Angular resolution is about 10 arc-second. The gray-scale bar is labelled in Jy units starting at the $4\sigma$ level of 17.2 µJy.

**Observations and interpretation**

The GRS 1758-258 radio jets were re-observed with the Jansky Very Large Array (VLA) of NRAO in 2016 March 04-22 at the 6 cm wavelength in the C-configuration of the array. The observing runs were carefully calibrated and combined with archival VLA data from 1992, 1993, 1997 and 2008 to produce the deepest radio view of this microquasar. With a 4.3 µJy noise level, the resulting image, shown in Fig. 1, noticeably improves previous sensitive observations (Martí et al. 2015).

GRS 1758-258 is the first microquasar where winged radio features have been detected. As sketched in Fig. 1, they are strongly similar to those observed in WRGs with Z-type morphology. The lengths of the Z wings are comparable to the full extension of the main jet flow (about 3 arc-minute), equivalent to a few pc at a distance of 8.5 kpc. The well-known scaling principles of fluid mechanics lead us to propose that, at least in some cases, the same physical mechanisms giving rise to this microquasar wings should also operate in the extragalactic context.

In GRS 1758-258, the spin-flip wing origin can certainly be ruled out. Being an X-ray binary system hosting a non-degenerate companion star, the possibility of a previous black hole merger is excluded from the evolutionary point of view. Only one black hole or compact object has ever been hosted by the system. Similarly, conical precession, or long-term changes in the accretion disc orientation, cannot be reconciled with the known properties of this microquasar. Indeed, they would require a very large cone angle or unrealistically long realignment time scales. The only physical scenario left to be consistent with the wing existence in this Galactic jet source is the one based on simple hydrodynamic backflow (similar as in Gopal-Krishna et al. 2003, 2012). The jet-medium interaction, required in such a backflow scenario, is also supported by the distribution of molecular gas in the source vicinity. Based on the Dame et al. (2001) survey of CO emission, a molecular cloud with kinematic distance similar to that of GRS 1758-258 appears as a likely collision target of the jets (Martí et al. 2017).

**Summary and conclusions**

The analogy of microquasars and other extragalactic jet sources has been expanded to include the WRG phenomenon thanks to the terminal jet regions of GRS 1758-258. While doing so, the hydrodynamic backflow scenario for the formation of winged features becomes strongly supported. Moreover, provided that backflow predominates, the coalescence of super-massive black holes in WRGs could be less frequent than previously considered. As a side implication, their contribution to the background of gravitational waves would need to be revised downwards.

**Acknowledgements:** The National Radio Astronomy Observatory is a facility of the National Science Foundation operated under cooperative agreement by Associated Universities, Inc. We also acknowledge funding support by the Agencia Estatal de Investigación grants AYA2016-76012-C3-1-P and AYA2016-76012-C3-3-P from the Spanish Ministerio de Economía y Competitividad (MINECO), by the Consejería de Economía, Innovación, Ciencia y Empleo of Junta de Andalucía under research group FQM-322, by grant MDM-2014-0369 of the ICCUB (Unidad de Excelencia 'María de Maeztu'), and by the Catalan DEC grant 2014 SGR 86, as well as FEDER funds. V.B.R. was also funded by MINECO and the European Social Funds within the Ramón y Cajal fellowship program, as well as the Marie Curie Career Integration Grant 321520.